# Effect of the bubble deformation in the3D nonlinear laser wake-field acceleration

Ershad Sadeghi Toosi[1], Saeed Mirzanejhad[2], Davoud Dorranian[1]

[1]Plasma Physics Research Center, Science and Research Branch, Islamic Azad University, Tehran, Iran

[2]Department of Atomic and Molecular Physics, Faculty of Basic Science, University of Mazandaran, Babolsar, Iran

Correspondence should be addressed to S. Mirzanejhad; Saeed@umz.ac.ir

**Abstract**

A new analytical approach for bubble deformation was used for optimization of the electron acceleration in the 3D highly nonlinear laser wake-field regime. Injection of the electron bunch with initial velocity in the bubble was considered in the inhomogeneous plasma with linearly density ramp. The researchers show that deformation of the bubble shape has an efficient role on the trapping of the electrons in the acceleration region. The influence of the linearly density ramp on the electron bunch trapping ratio and its mean energy was considered by the numerical method.

# 1. Introduction

Laser wake-field acceleration of electrons is a new efficient practical acceleration scheme which is investigated in several laboratories for a few decades[1-4]. As the laser propagates into the plasma, its front edge diffracts while the rest of the pulse self-focuses. Electrons are expelled by the ponderomotive force from the high-intensity volume and a plasma cavity which is void of electrons is produce behind the laser pulse. The expelled electrons travel along the field lines and accumulate at the back of the cavity where they can be injected and accelerated. This regime is also called the bubble regime[5,6]. The laser strength $a_0$ is mainly used to distinguish between linear and non-linear regimes of laser wake-field excitation that the bubble regime occurs for $a_0 \gg 1$[7].

Somelimitations on the electron energy gain in the laser wake-field acceleration are injection, trapping, dephasing, beam quality and the effects produced by the electromagnetic field of the accelerated electron bunch[8,9]. Injection or trapping of background electrons into the acceleration region occur by internal or external drivers.When the electron is accelerated to high energy, the electron can go faster than the plasma wave and leave the acceleration region, which is called dephasing. The dephasing length, which is determined by the plasma density, limits the acceleration length in LWFA. The dephasing length increases when the electron density decreases[10, 11].

Several optical plasma electron injection schemes in wake-field acceleration have been proposed to enhance the energetic electron population, such as trapping by colliding pulses, an intense laser beam interacts with a slowly varying downward density transition, a short and intense laser pulse passing across a parabolic plasma channel transversely, ionization-induced trapping and tightly focused intense laser beam[12-26]. The increase in the electron energy by increasing the acceleration length has already been demonstrated using a capillary waveguide[27], but another simple method to increase the electron energy is effect of the plasma density ramp which is easily generated by using the supersonic gas nozzle[28]. A transient density ramp can be produced by a laser pre-pulse or in specially designed gas targets [29, 30]. The electron injection in thedescending plasma density target was studied in detail in experiments which were presented in Geddes *et al.*[31].

For matched laser–plasma parameters, the electron acceleration can occur in an unlimited acceleration regime. This regime implies that the electron inside the wake wave does not leave the acceleration phase, i.e. the dephasing length is formally infinite, and its energy grows in time, being not limited asymptotically. This condition can be realized when the laser pulse generates the wake wave in a down-ramp density plasma within the framework of the 1D approximation, assuming that the wake wave is generated by an ultra-short laser pulse with constant amplitude $a_{las} < 1$, which has been shown in Bulanov *et al.* that in the plasma target with a downward gradient density, $n \sim x^{-2/3}$, the electron energy grows proportionally to $\sim x^{1/3}$[32-36].

Kim *et al.* proposed the effect of the density upward and the density downward ramp structures on the electron energy. The downward density ramp reduces the electron energy due to the lag of the acceleration region. The upward density ramp can increases the electron energy because the acceleration region is fast, which effectively increases the dephasinglength[37,38]. Suk *et al.* also proposed a scheme for plasma electron trapping by using a density transition[39].

A new analytical formalism introduced by Toosi *et al.*[40], were used for study the effect of the bubble deformation in LWFA. Analytical equations for bubble formation behind the laser pulse and the relationship between the longitudinal and transverse bubble radiuses, $R_z$ and $R_r$ are found as a function of normalized maximum laser pulse amplitude $a_0$, pulse length $l_p$, laser beam waist $w_0$ and its normalized frequency $\omega/\omega_p$ were borrowed from Toosi *et al* [40].

In this manuscript, effect of bubble deformation on the trapping and acceleration of electrons in the inhomogeneous plasma is considered. External injection of electron bunch with initial energy recognize in the numerical analysis. In section II, brief review of analytical formalism for bubble formation is mentioned. Section III, includes numerical results for trapping and acceleration of electron bunch in the bubble regime of the LWFA.

## 2. Analytical Formalism

In the bubble regime of the laser wake-field acceleration, the bubble structure passes through the background plasma with the group velocity of the laser pulse. In the equilibrium (quasi-static) state, some background electrons pass through the bubble but others scattered by the ponderomotive force of the laser pulse and lay in the electron sheath layer around the bubble boundaries. These electrons introduce an electron sheath around the bubble which shield electrostatic field of the bubble around the bubble boundaries[40]. Assume the Gaussian laser pulse propagates along the z-axis with amplitude,

$$A_0(x,y,z,t) = a_0 \frac{w_0}{w(z)} \exp[-\frac{r^2}{w^2(z)}] \exp[-\frac{4(\zeta-\zeta_0)^2}{lp^2}] \quad , \tag{1}$$

Where

$$\zeta = z - v_g t, w(z) = w_0 \sqrt{1 + \frac{4(z-z_{0g})^2}{k^2 w_0^4}}, \quad r^2 = x^2 + y^2, \quad \zeta_0 = z_{0l}, \quad z_{0l} = z_{0g} = z_0 = 0.$$

In the numerical results, The researchers select $w_0 = 6 \, \mu m$ for the laser beam waist, $\tau_l = 20 \, fs$ for the pulse duration, the wave length $\lambda_l = 1 \mu m$ and the laser normalized vector potential is

$a_0 = eE_0/mc\omega = 10$, which corresponds to the laser intensity, $I \approx 1.4 \times 10^{20} \frac{W}{cm^2}$. The normalized longitudinal and transverse components of the laser pulse ponderomotive force are as follows,

$$F_{pond}{}^r = -\frac{\partial}{\partial r}V_{pond} = -\frac{|A|}{2\sqrt{1+\frac{|A|^2}{2}}}\frac{\partial|A|}{\partial r} = \frac{|A|^2 r}{w^2(z)\beta_0\sqrt{1+\frac{|A|^2}{2}}},$$

$$F_{pond}{}^z = -\frac{\partial}{\partial z}V_{pond} = -\frac{|A|}{2\sqrt{1+\frac{|A|^2}{2}}}\frac{\partial|A|}{\partial z} = \frac{2|A|^2}{\beta_0\sqrt{1+\frac{|A|^2}{2}}}[\frac{(z-z_0g)}{k^2 w_0{}^2 w^2(z)}(1-\frac{2r^2}{w^2(z)})+\frac{2(\zeta-\zeta_0)}{lp^2}], \quad (2)$$

And the normalized longitudinal and transverse electromagnetic force of the bubble as follows,

$$F_{bub}{}^r = -\frac{\beta_0{}^2 r}{2(1+\beta_0)}, \quad F_{bub}{}^z = -\frac{\beta_0 \xi}{1+\beta_0}. \quad (3)$$

In the previous work, the dimensions of the ellipsoid bubble are obtained as a function of the laser pulse normalized parameters $a_0$ (amplitude), $\widetilde{w_0} = k_p w_0$ (waist), $\widetilde{l_0} = k_p l_0$ (length) and normalized frequency $\omega/\omega_p$ [40]. Geometrical structure of the bubble boundary was obtained by balancing between ponderomotive and bubble forces (Eqs. 2 and 3). Parameters bubble dimensions used in this study were borrowed from Toosi *et al* [40],

$$R_z\left(a(z), l_p, \frac{\omega}{\omega_p(z)}\right) = -1.1834 + 0.6508\, a(z) + 0.465 l_p + 0.688\left(\frac{\omega}{\omega_p(z)}\right)^{-3.19},$$

$$R_r\left(a(z), l_p, w(z), \frac{\omega}{\omega_p(z)}\right) = -32.44 - 0.021 a^2(z) + 0.8622\, a(z) - 0.9041 w^2(z) +$$

$$10.08\, w(z) + 3.547\left(\frac{\omega}{\omega_p(z)}\right)^{-2.319} + 21.85\, l_p^{-1.743}, \quad (4)$$

In which ~ disappears from the normalized parameters, $a(z) = a_0 \frac{w_0}{w(z)}$ and $\omega_p(z)$ is plasma local frequency for the inhomogeneous plasma. For example, for the inhomogeneous plasma with linear ramp, $\omega_p(z) = \omega_{p0}(1 + \beta z)$, where $\omega_{p0}$ is a plasma frequency at z=0 and $\beta$ is a plasma density ramp parameter. On the other hand, equation (4) can be written in the following multiplicative form,

$$R_z\left(a(z), l_p, \frac{\omega}{\omega_p(z)}\right) = -1.1834 + 0.6508\, a_0 \frac{w_0}{w(z)} + 0.465 l_p + 0.688 \left(\frac{\omega}{\omega_{p0}(1+\beta z)}\right)^{-3.19},$$

$$R_r\left(a(z), l_p, w(z), \frac{\omega}{\omega_p(z)}\right) = -32.44 - 0.021 a_0^2 \frac{w_0^2}{w^2(z)} + 0.8622\, a_0 \frac{w_0}{w(z)} - 0.9041 w^2(z) +$$
$$10.08\, w(z) + 3.547 \left(\frac{\omega}{\omega_{p0}(1+\beta z)}\right)^{-2.319} + 21.85\, l_p^{-1.743}, \quad (5)$$

When the laser pulse propagates through the plasma, bubble dimensions varies with variation of the beam waist w(z) and pulse amplitude a(z). The bubble deformation in five different position of the laser pulse ($-z_R < z_{pulse} < z_R$) with dimensionless parameters $a_0 = 10, w_0 = 3.6, l_p = 2$ in the homogenous plasma with $\omega/\omega_p = 10$ are shown in figure (1).

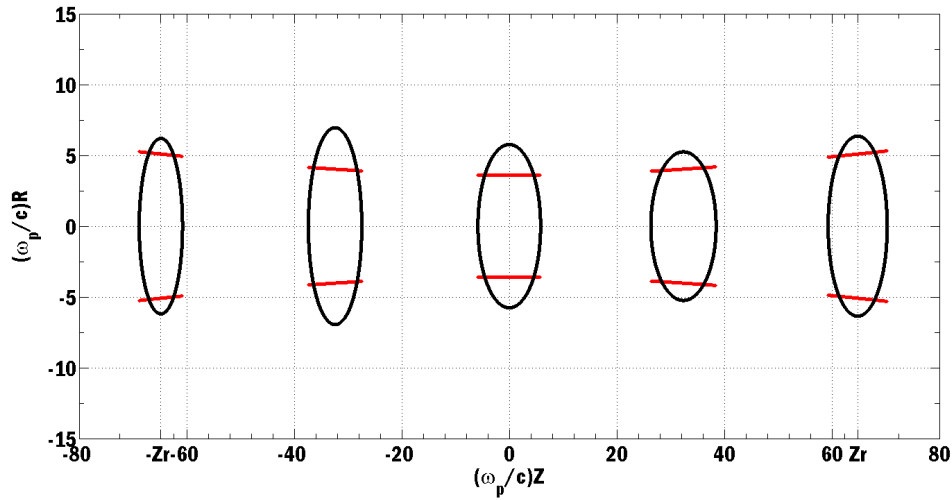

Fig.1. Bubble deformation at different positions ($-z_R < z_{pulse} < z_R$).

In the next section, injection of the electron bunch in front of the bubble is considered by numerical simulation. The researchers show that variation of the bubble dimensions due to the diffraction of Gaussian laser pulse or refraction in the inhomogeneous plasma have an important role to the trapping and acceleration of electrons.

## 3. Simulation Results

In this section, numerical results are obtained for an electron bunch with $10^5$ electrons which is distributed randomly with longitudinal and transverse dimensions σz=σr=λ=1μm, respectively. This electron bunch is injected in front of the laser pulse which is identified in previous section (Fig. 2). Initial velocity components of electrons are assigned according to the Boltzmann distribution with different longitudinal and transverse spreads. Electrons average initial energy is in the range, $\langle K_0 \rangle$ =0-2MeV with 0.1% longitudinal energy spreadandbunch transverse emmitance is 1 mm mrad. These parameters simulate a bunch with $3 \times 10^{16} \frac{1}{cm^3}$ number density and about 16 nC charge. In the second part of this section,inhomogeneous plasma with linear ramp is used to optimize acceleration characteristics.The average final electron kineticenergy,$\langle K_{bunch} \rangle$, ratio of trapped and accelerated electronsand bunch quality are presented for different initialparameters in figures (3-6).

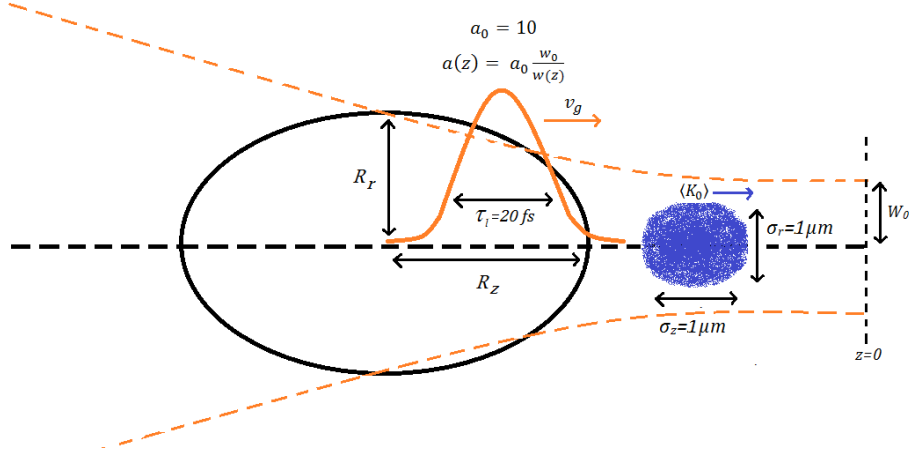

Fig. 2. Schematic diagram and physical parameters of the electron bunch injection in front of the laser pulse.

The initial position of electrons with respect to the origin (laser pulse focus) is an important parameter in acceleration mechanism. Proper synchronization between the bunch and the laser field can be adjusted for an efficient acceleration gradient. Figure (3) shows electrons final energy in terms of the interaction position. The interaction position obtains by the following approximate formula,

$$Z_{int} \approx Z_{0l} + (Z_{0b} - Z_{0l})/(1 - \beta_{0b}) \qquad (6)$$

In which $Z_{0l}$ is initial position of the laser pulse and $Z_{0b}$ and $\beta_{0b}$ are initial position and velocity of the electron bunch respectively. In this figure, initially electron bunch is injected toward the bubble with 0.5 MeV energy.

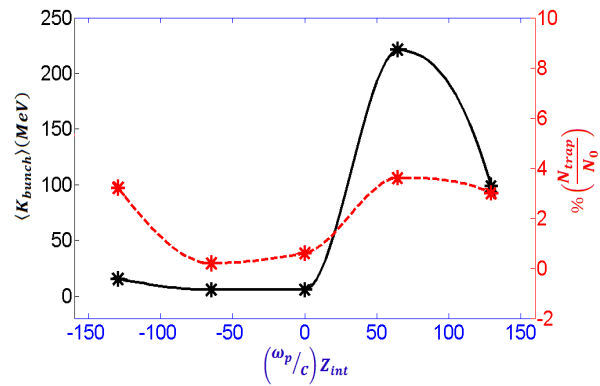

Fig. 3, Electron bunch final average energy with respect to the interaction position, $z_{int}$ that the initial bunch kinetic energy is 0.5 MeV.

Figure (3) shows the importance of the synchronization between laser pulse and electron injection. Efficient acceleration reaches if interaction occurs after the focus point. Optimum location is about one Rayleigh length after focus, where electron bunch energy grows to 220 MeV. According to the figure 1 at this position,bubble has largest dimensions, which leads to largest dephasing length and largest trapping ratio. At this point about 4% of electrons trapped and accelerated in the bubble.

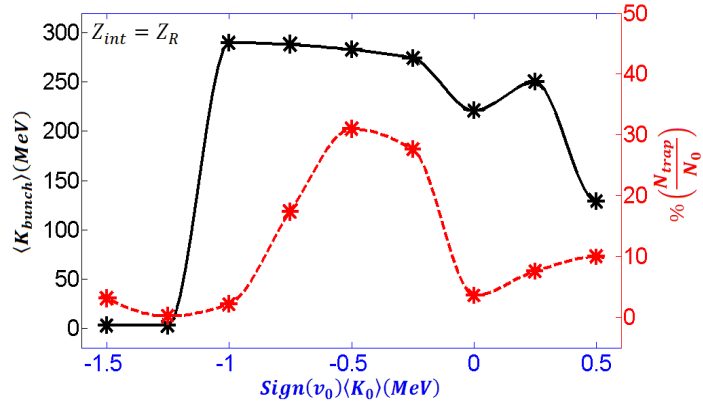

Fig.4.Electron bunch average energy with respect to the different initial bunch kinetic energy.

Figure (4) shows significance of the initial bunch kinetic energy and its injection direction in the acceleration process. Injection in the counter-propagation of the laser pulse and bubble has surprising results. Background electrons gain 220 MeVenergy,but electron bunch with initial energy in the range (0.25-1MeV)and in opposed to the bubble direction, achieve280MeVkinetic energy. As well as, trapping ratio in this case grows from 4% to greater than 30%. This result suggests advantage of the counter-propagation injection procedure for the researchers selected parameters.

In the last part of this section, the researchers focus on the effect of plasma density ramp on the results of electron trapping and acceleration in the optimum condition. It was mentioned that negative density ramp increasesdephasing length and energy gain. The researchers choose linear density ramp with ramp parameter $\beta$ as follow,

$$\omega_p(z) = \omega_{p0}(1 + \beta z) \tag{7}$$

Figure (5) shows average accelerated electrons energy and trapping rate for different ramp parameters $(-0.1 < (c/\omega_{p0})\beta < 0.1)$ for the optimum conditions, $\langle K_0 \rangle = 0.5$ MeV and $Z_{int} = Z_R$. It is clear that negative ramp $((c/\omega_{p0})\beta < -0.05)$ grows energy gain from 280 MeV to about 380 MeV. Likewise trapping ratio grows from 20% to about 40% in the negative density ramp situation. Other salience result in this figure is large electron trap ratio >70% in the positive ramp parameter, $(c/\omega_{p0})\beta = 0.03$.

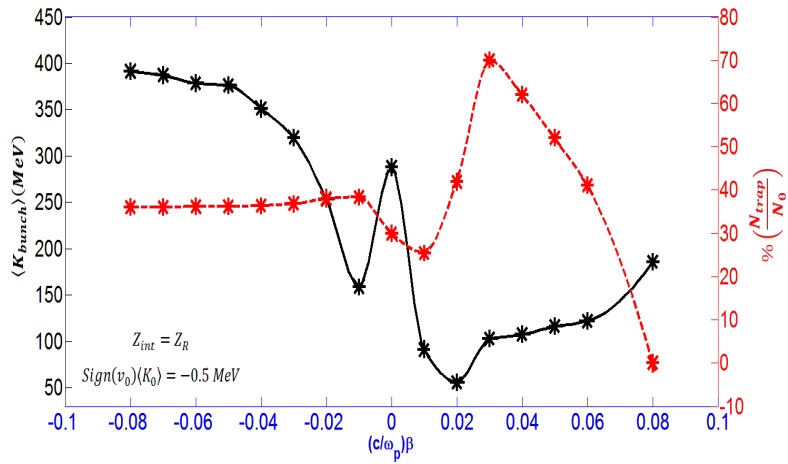

Fig.5. Electron bunch final energy(black solid line)and electron trapping ratio (red dashed line) with respect to different ramp parameter $\beta$, ($\langle K_0 \rangle = 0.5$ MeV, $Z_{int} = Z_R$ ).

In figure (6), interaction of bubble with electron bunch is shown in for different interaction time. Trapping, focusing and acceleration of electrons in the rear side of the bubble are shown in these figures.

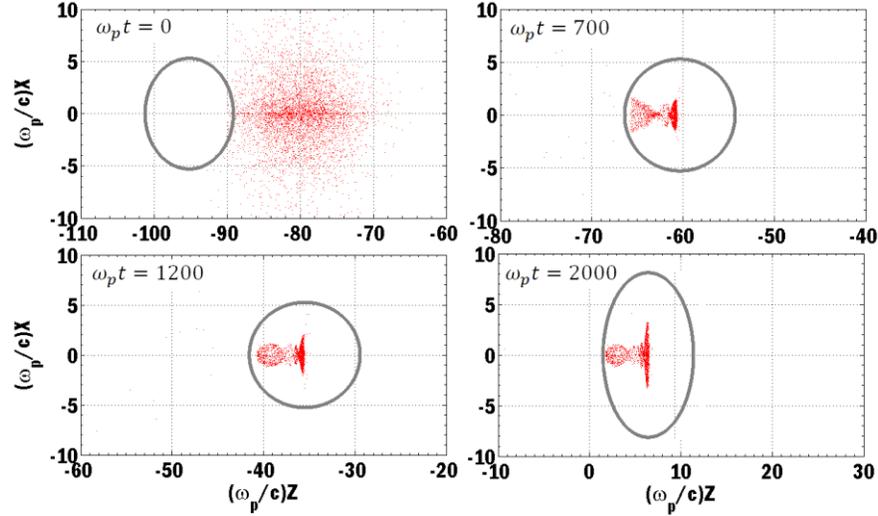

Fig.6. Electron density profile at $\omega_p t = 0, 700, 1200, 2000$.

## 4. Conclusion

Electron acceleration in the highly nonlinear regime of the 3D laser wake field acceleration scheme is interested for many researchers. One of the imperfections of this acceleration regime is low charge of the accelerated electron in the self-injected condition. Some injection methods such as, optical injection, plasma density ramp or pre-accelerated bunch injection were proposed for omission of this imperfection. In this paper, the researchers regarded to the external injection in the inhomogeneous plasma. The researchers showed that the synchronization between laser pulse and injected electron bunch has serious effect on the acceleration results. Efficient acceleration reaches if interaction occurs after the focus point. Optimum location is about one Rayleigh length after focus, where electron bunch energy grows to up to the 220 MeV. At this point about 4% of electrons were trapped and accelerated in the bubble. Significance of the initial bunch kinetic energy and its injection direction in the acceleration process are shown in numerical results. Injection in the counter-propagation of the laser pulse has surprising results. Background electrons gain 220 MeV energy, but electron bunch with initial energy in the range (0.25-1 MeV) and in opposed to the bubble direction, achieve 280 MeV kinetic energy. As well as, trapping ratio in this case grows from 4% to greater than 30%. This result suggests the advantage of the counter-propagation injection procedure for the researchers selected parameters. Finally, numerical results for the inhomogeneous plasma show that negative ramp $((c/\omega_{p0})\beta < -0.05)$ grows energy gain from 280 MeV to about 380 MeV. Likewise, trapping ratio grows from 20% to about 40% in the negative density ramp situation.

These results may be used for increasing energy gradient and charge of the electron bunch in the bubble regime of the LWFA. Counter-propagation of the injected electron bunch is a new proposed scheme which has surprising results in laser-plasma acceleration laboratories.

**Conflicts of Interest**

The authors declare that there are no conflicts of interest regarding the publication of this paper.